\documentclass{PoS}

\title{Radio-bimodality in galaxy clusters and
future observations at
low radio frequencies: constraining the origin of giant 
radio halos}

\ShortTitle{Radio halos and future low frequency radio observations}

\author{\speaker{Gianfranco Brunetti}\\
        INAF - Istituto di Radioastronomia, via P. Gobetti 101, 40129 Bologna, Italy\\
        E-mail: \email{brunetti@ira.inaf.it}}
\author{Rossella Cassano\\
        INAF - Istituto di Radioastronomia, via P. Gobetti 101, 40129 Bologna, Italy\\
        E-mail: \email{rcassano@ira.inaf.it}}

\abstract{Radio observations discovered large scale non thermal sources in the 
central Mpc regions of dynamically disturbed galaxy clusters (radio halos).
The morphological and spectral properties of these sources suggest that 
the emitting electrons are accelerated by spatially distributed and gentle 
mechanisms, providing some indirect evidence for turbulent acceleration in the 
inter-galactic-medium (IGM).
Radio and X-ray surveys allow to investigate the statistics of
radio halos and unveil a bimodal behaviour of the radio 
properties of galaxy clusters: merging clusters host radio halos and trace
the well known radio--X correlation, while more relaxed clusters 
do not host radio halos and populate a region well separated from that 
spanned by the above correlation.
This appears consistent with the hypothesis that relativistic electrons
are reaccelerated by MHD turbulence generated during cluster mergers.
In the context of this model the population of radio halos consists of
a mixture of halos with different spectral propertis, most of them with
very steep spectrum and visible only at low radio frequencies.
For this reason the future LOFAR surveys may provide a robust test
to this theoretical hypothesis.}

\FullConference{ISKAF2010 Science Meeting - ISKAF2010\\
		June 10-14, 2010\\
		Assen, the Netherlands}

\begin{document}

\section{Introduction}

Radio observations of galaxy clusters prove the presence of 
non-thermal components, magnetic fields and relativistic particles, 
mixed with the hot Inter-Galactic-Medium (IGM) (e.g., Ferrari et al., 2008).
These components deserve some 
attention since they provide an additional source of energy in the IGM 
and may drive still unexplored physical processes modifying our simplified 
view of the IGM (Schekochihin et al. 2005; Subramanian et al. 2006; 
Brunetti \& Lazarian 2007; Guo et al. 2008).

During cluster mergers a fraction of the gravitational binding--energy 
of Dark Matter halos that is converted into internal energy of the 
barionic matter
can be channelled into the amplification of the magnetic fields
(e.g. Dolag et al. 2002; Subramanian et al. 2006; Ryu et al. 2008) and into
the acceleration of particles via shocks and turbulence 
(e.g. Ensslin et al 1998; Sarazin 1999; 
Blasi 2001; Brunetti et al. 2001, 2004; Petrosian 2001; Miniati et al. 2001; 
Ryu et al. 2003; Hoeft \& Bruggen 2007; 
Brunetti \& Lazarian 2007; Pfrommer 2008).

\noindent
Upper limits to the gamma ray emission from galaxy clusters,
obtained by FERMI and Cherenkov telescopes (eg., Aharonian et al
2009a,b; Aleksic et al 2010; Ackermann et al 2010), together with 
constraints from complementary approaches based on radio observations
(eg., Reimer et al 2004; Brunetti et al 2007) 
suggest that relativistic protons contribute to less than a few percent
of the energy of the IGM, at least in the central Mpc--sized regions.

\noindent
On the other hand,
the acceleration of relativistic electrons in the IGM 
is directly probed by radio 
observations of diffuse synchrotron radiation from galaxy clusters.
{\it Radio halos} are the most spectacular examples of cluster-scale radio sources, 
they are diffuse radio sources that extend on
Mpc-scales in the cluster central regions and are found in about $1/3$ of 
massive galaxy clusters with complex dynamics
(eg. Feretti 2002; Ferrari et al. 2008; Cassano 2009).

\noindent 
The origin of radio halos is still debated.
According to a first hypothesis, radio halos are due to 
synchrotron emission from secondary electrons generated by p-p collisions
(Dennison 1980;
Blasi \& Colafrancesco 1999; Pfrommer \& Ensslin 2004), in
which case clusters are (unavoidably) gamma ray emitters due to the decay of
the $\pi^o$ produced by the same collisions.
The non-detections of nearby galaxy clusters at GeV energies 
by FERMI
significantly constrain the role of secondary electrons in the non-thermal 
emission (Ackermann et al. 2010). 
Most important, the spectral and morphological properties of a 
number of well studied
radio halos appear inconsistent with a pure hadronic origin 
of the emitting
particles (eg. Brunetti et al 2008, 2009; Donnert et al 2010a,b; 
Macario et al 2010).

\noindent
A second hypothesis is based
on turbulent reacceleration of relativistic particles in 
connection with cluster--mergers events (eg.,
Brunetti et al. 2001; Petrosian 2001; 
Fujita et al 2003; Cassano \& Brunetti 2005).
The acceleration of thermal electrons to relativistic energies by 
MHD turbulence in the IGM faces serious drawbacks based on energy 
arguments (eg., Petrosian \& East 2008), consequently in these models 
it must be assumed a pre-existing population of relativistic electrons in
the cluster volume that provides the seed particles to reaccelerate during
mergers.

In addition to the spectral and morphological properties of radio halos
and to the constraints on the gamma ray emission from clusters, 
the statistical
properties of radio halos and their connection with cluster formation
and evolution provide crucial insights into the physics and origin of
non-thermal components in the IGM.

\noindent
In this paper we focus on the most recent advances in the study of the
statistical propertis of radio halos and their connection with cluster
mergers (radio bimodality), and discuss the importance of future surveys
at low radio frequencies to test present models.

\section{The radio bimodality of galaxy clusters}

A first observational settlement of the properties of radio halos, 
at $z\leq 0.2$, and of their connection with cluster mergers
has been obtained by means of deep follow ups with the VLA 
of candidates radio halos identified with the NVSS and WENSS radio 
surveys (Giovannini et al., 1999; Kempner \& Sarazin 2001). 
A step forward has been recently achieved through a 
project carried out with the Giant Metrewave Radio 
Telescope (GMRT, Pune-India) at 610 MHz, 
the ``GMRT Radio Halo Survey'' (Venturi et al. 2007, 2008).

\begin{figure*}[t]
\begin{center}
\includegraphics[width=0.52\textwidth,bb=0 50 670 800,clip
]{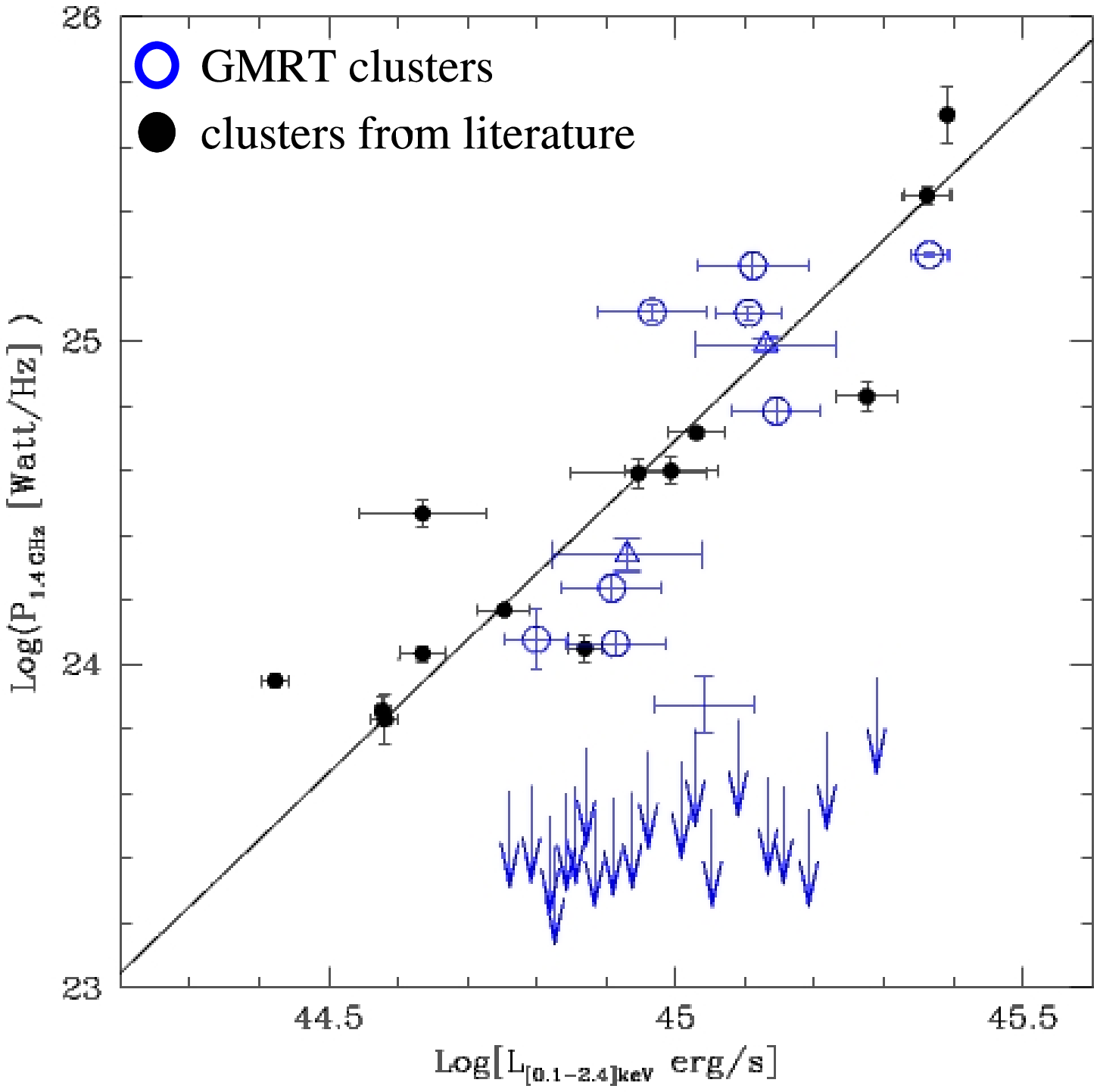}
\hskip -1cm
\includegraphics[width=0.5\textwidth,bb=0 50 850 999,clip
]{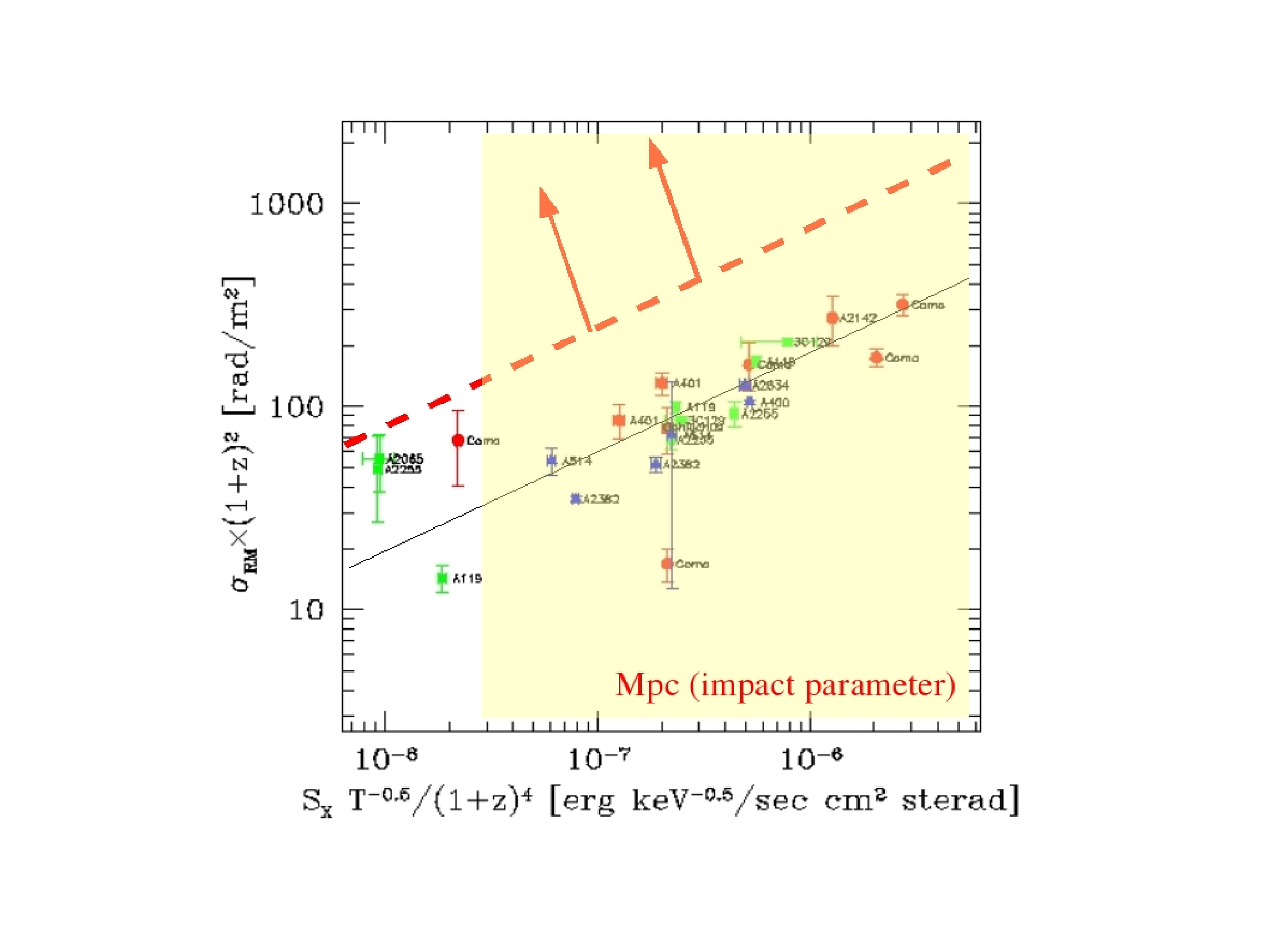}
\caption[]{{\bf Left Panel}: distribution of galaxy clusters
in the radio -- X-ray luminosity diagram (from Brunetti et al 2009).
Blue symbols mark clusters of the GMRT sample.
The solid line is the radio -- X-ray correlation of radio halos.
{\bf Right Panel}: distribution of clusters (several line of sights
per cluster) in the $\sigma_{RM}$--X-ray brightness diagram (from
Govoni et al 2010).
Solid thin line marks the case where all clusters have the same
magnetic field properties, the thick dashed line mimics the effect
of a magnetic field 3 times larger.
Cluster sources at (about) $\leq$ Mpc (projected) distances from cluster
centers fall in the shadowed region.} 
\label{fig:Lr_Lx}
\end{center}
\end{figure*}

The ``GMRT Radio Halo Survey'' allows to unveil a 
{\it bi-modal} behaviour of the radio properties of galaxy 
clusters (Fig.~1a), with radio-halo 
clusters and clusters without radio halos clearly separated 
(Brunetti et al.~2007).
The radio bimodality sheads new light on the evolution of 
non-thermal components in galaxy clusters and on their 
connection with cluster dynamics.
More recently Cassano et al.(2010a) found that the bimodal 
behaviour in Fig.1a has a correspondence in terms of
cluster dynamical properties, with radio halos found in
merging clusters and ``radio quiet'' clusters being systematically 
more relaxed systems.

\noindent
This suggests the following 
coupled evolution between radio halos and cluster
dynamics :

\begin{itemize}
\item{\it i)} galaxy clusters host giant radio halos for a period of time,
in connection with cluster mergers, and populate the 
$P_{1.4}$--$L_X$ correlation (Fig.1a);

\item{\it ii)} at later times,
when clusters become dynamically relaxed, the Mpc-scale
synchrotron emission 
is gradually suppressed and clusters populate 
the region of the upper limits.
\end{itemize}

When restricting to clusters of the GMRT complete sample, 
Fig. 1a provides a fair statistical sampling of the 
evolutionary
flow of X--ray luminous clusters in the $P_{1.4}$--$L_X$ plane
at z=0.2-0.4.
Radio-halo clusters in the GMRT sample, that are always dynamically
disturbed systems, 
must be the ``youngest'' systems, where an ongoing
merger, leading to their formation (or accretion of a sizable fraction
of their mass), is still supplying energy to maintain the synchrotron
emission.
On the other hand, clusters with radio upper limits, 
that are more relaxed than radio halo clusters,
must have experienced the last merger at earlier epochs : after the last
merger they already had sufficient time for suppression of the
synchrotron emission and consequently
they should be the ``oldest'' systems in the GMRT sample.
Clusters in the ``empty'' region may be ``intermediate'' systems 
at late merging phases, where synchrotron emission is being suppressed,
or ``very young'' systems in the very early phases of a merging
activity, where synchrotron emission is increasing.

19 clusters in the GMRT sample have 
$L_X \geq 8.5 \times 10^{44}$erg s$^{-1}$,
in which case the radio power of giant radio halos is $\sim$1 
order of magnitude
larger than the level of the
radio upper limits. Of these 19 clusters 5 host
giant radio halos (and 2 mini-halos in cool-core clusters),
11 clusters are ``radio quiet'' and only RXJ1314 is in the
``empty'' region.
This allows for estimating the life--time of radio halos, $\tau_{rh}\approx
1$ Gyr, and the time clusters spend in the ``radio--quiet'' phase, 
$\tau_{rq} \approx 2-2.5$ Gyr (Brunetti et al 2009).
Most importantly, at these luminosities,  
the ``emptiness'' of the region between radio halos and ``radio quiet'' 
clusters in the $P_{1.4}$--$L_X$ diagram 
constrains the time-scale of the evolution (suppression and amplification) 
of the synchrotron emission in these clusters (Brunetti et al. 2007, 2009).
The significant lack of clusters in this region suggests
that this time-scale is much shorter than both the ``life-time''
of clusters in the sample and the period of time clusters spend in the
radio halo stage.
Monte Carlo analysis of the distribution of clusters in Fig. 1a shows that 
the time interval that clusters spend in the ``empty'' region
(thus the corresponding time--scale for amplification and suppression
of radio halos) is $\tau_{evol} \approx 200$ Myr, 
with the probability that $\tau_{evol}$ is as large as 1 Gyr 
$\leq 1$\% (Brunetti et al. 2009).

\begin{figure*}[t]
\begin{center}
\includegraphics[width=0.5\textwidth
]{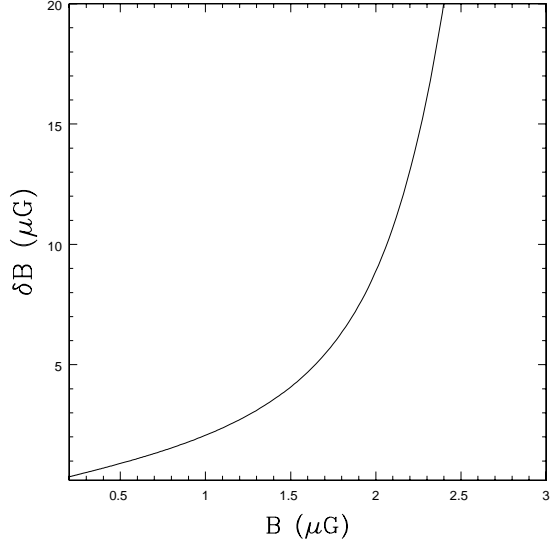}
\caption[]{The minimum excess of magnetic field strength that is necessary 
to have a synchrotron amplification of about 10 times in radio
halo clusters (at redshift 0.25)
is shown as a function of the magnetic field in ``radio quiet'' clusters.} 
\label{fig:Lr_Lx}
\end{center}
\end{figure*}

\section{Implications of bimodality 
for the origin of giant radio halos}

The radio properties of galaxy clusters
are driven by the evolution of the relativistic components 
(B and particles) in the IGM.
The tight constraints on the time--scale evolution of the
clusters radio properties, $\tau_{evol} \approx 200$ Myr, 
provides crucial information on the
physics of the particle acceleration and 
magnetic field amplification.

\subsection{Magnetic field evolution ?}

A first possibility to explain the bimodality
is that cluster mergers amplify 
the magnetic field in the IGM leading to the amplification of
the synchrotron emission on Mpc scales.
In this case, merging clusters hosting
radio halos have larger magnetic fields, $\delta B + B$, 
with the excess $\delta B$ being generated during mergers and then 
dissipated when clusters become ``radio quiet'' and 
more
dynamically relaxed (Brunetti et al.~2007, 2009; Kushnir et
al.~2009)\footnote{see also Keshet \& Loeb arXiv:1003.1133}.

\noindent
The condition for a suppression $\geq$10 in terms of synchrotron
emission (Fig. 1a) constrains the ratio between the magnetic 
fields in radio halos, $B + \delta B$, and that in ``radio quiet'' 
clusters, $B$ :

\begin{equation}
( {{ B + \delta B }\over{ B }} )^{\alpha-1}
{{ 1 + ( {{B_{cmb} }\over{B }} )^2 }\over
{1 + ( {{B_{cmb} }\over{B + \delta B}} )^2 }}
\geq 10
\label{eq:gap}
\end{equation}

\noindent
where $B_{cmb} =3.2 (1+z)^2 \, \mu$G is the equivalent field of the CMB
and $\alpha \sim 1.3$ is the typical
synchrotron 
spectral index of radio halos. 
The amplification factor of the magnetic field 
is shown in Fig. 2 for $z\approx 0.25$, typical of
GMRT clusters.
In the case $B + \delta B << B_{cmb}$, 
the energy density of the magnetic field in radio halo 
clusters should be $\geq$10 times larger than that
in "radio quiet" clusters, and even larger ratios
must be admitted if $B + \delta B >> B_{cmb}$.

This significant difference between the magnetic field strength
in radio halo and "radio quiet" clusters disfavours an interpretation
of the radio bimodality based on the amplification/suppression of
the cluster magnetic field.
Indeed Faraday Rotation measurements (RM) in galaxy
clusters do not find any statistical difference between
the energy density of the large scale (10-100 kpc coherent scales) magnetic 
field in radio halo clusters and that in "radio quiet" clusters 
(e.g., Carilli \& Taylor 2002).
For example, in a recent paper, Govoni et al (2010) presented a study
of RM of radio sources in a sample of hot galaxy clusters, 
including both ``radio-quiet'' and clusters
with well known radio halos (eg. Coma, A2255, ..).
In Fig. 1b 
we show the $\sigma_{RM}-S_X$ distribution from Govoni et al.,
$\sigma_{RM}$ and $S_X$ being the $\sigma$ of the RM and the X--ray
(thermal) cluster brightness measured along several 
radio sources at different (projected) distances from cluster centers.
Since $\sigma_{RM} \propto \Lambda_c \int (n_{th} B_{\Vert})^2 dl$
($\Lambda_c$ the field coherent scale)
Fig.1b provides an efficient way to separate the magnetic and thermal
properties of clusters.
Although the still poor statistics, 
Fig.1b shows that the magnetic field strength in
``radio--quiet'' and radio halo clusters is similar (the thin solid
line is obtained for a fixed value of $B$ and $\Lambda_c$).
On the other hand, the dashed line in Fig.1b marks 
the region where radio halo
clusters should have been found if the radio bimodality is
driven by the magnetic field 
amplification/suppression (according to Fig. 2).

\subsection{Bimodality and acceleration/cooling 
of relativistic particles}

{\it RM suggests that a ``single'' mode exists for the magnetic
field in galaxy clusters and consequently 
that the observed radio bimodality implies 
a corresponding bimodality in terms of the
emitting relativistic particles in the IGM.}

\subsubsection{The case of pure secondary models}

Theoretically relativistic protons are expected to be the dominant
non-thermal particles component since they have long life-times and
remain confined within galaxy clusters
for a Hubble time (V\"{o}lk et al. 1996; Berezinsky, Blasi \& Ptuskin 1997;
Ensslin et al 1998).
The confinment of cosmic rays is a natural consequence of the Mpc
sizes and magnetization of galaxy clusters.
Assuming a Kolmogorov spectrum of the magnetic field fluctuations,
the time necessary to diffuse on Mpc scale is (eg Blasi \& Colafrancesco
1999) :

\begin{equation}
\tau_{diff} (Gyr) \approx 
65 \, R_{Mpc}^2 ({{E}\over{100 GeV}})^{-1/3}
B_{\mu G}^{1/3}
({{\Lambda_c}\over{20 kpc}})^{-2/3}
\label{taudiff}
\end{equation}

Tangled magnetic fields with coherent scales 
$\Lambda_c \approx 10-30$ kpc are
routinely derived through RM analysis of extended radio sources at
different (projected) distances from cluster centres in both
relaxed ("radio quiet") clusters and in radio halo clusters (eg.
Clarke et al 2001; Murgia et al 2004; Govoni et al 2010) implying 
diffusion time--scales of cosmic rays of many Gyrs.

\noindent
Consequently, classical hadronic models,  
where the emitting electrons are continuously generated
by p-p collisions, predict a population of secondaries that is
almost independent of the dynamical status
of the hosting clusters (eg., Blasi et al 2007 for review).
As a matter of fact suppression of radio halos on time--scales 
of $\sim$200 Myrs via (ad hoc) diffusion of cosmic rays from the 
central Mpc regions
implies extreme diffusion velocities, $\approx$ 100$\times v_A$,  
in which case plasma instabilities (eg streaming instability) are expected
to damp the diffusion process itself\footnote{After this paper was accepted
Ensslin et al.(arXiv:1008.4717) discussed the role of diffusion {\it
assuming} the idealized picture where cosmic rays can efficiently stream 
in relaxed (non turbulent) clusters. 
As mentioned above, an important aspect here is the generation of streaming 
instability and its effect in quenching cosmic ray streaming (due to 
scattering with MHD waves) under viable (less--idealized) physical conditions.
We will discuss this issue (in the {\it high--beta} ICM) in 
a forthcoming paper.}

\noindent
Thus the observed radio bimodality and the short $\tau_{evol}$ would
be difficult to reconcile with these models.

\subsubsection{The case of turbulent acceleration}

The radio bimodality and the constrained time--scale for the evolution
of clusters radio properties suggest that relativistic electrons
are accelerated in situ
on Mpc--scales (and maintained) during cluster mergers
and that they cool as soon as clusters become more relaxed.
Remarkably the cooling time of GeV electrons in the IGM is
$\sim 10^8$ yrs, much shorther than all the other relevant time--scales,
and may potentially fit the short value of $\tau_{evol}$ as 
constrained by the distribution of galaxy clusters in Fig. 1a.
 
As soon as large scale turbulence in the ICM reaches smaller, resonant, 
scales (via cascading or induced plasma instabilities, e.g. 
Brunetti et al. 2004, Lazarian \& Beresnyak 2006, Brunetti \& Lazarian 2007),
particles are accelerated and generate synchrotron emission.
In the case of radio halos emitting at GHz frequencies 
the acceleration process should be relatively efficient and 
particles get accelerated to the energies necessary to produce synchrotron
GHz--emission within a time-scale smaller than a couple of cooling times of 
these electrons, that is $\approx 100$ Myrs.
Although the large uncertainties in the way large scale
turbulence is generated in the IGM during cluster mergers, it is likely 
that the process persists for a few crossing times of the cluster-core 
regions, that is fairly consistent with a radio halo life-time $\tau_{rh}
\sim 1$ Gyr as derived in Section 3.1.

\noindent
Most importantly, the cooling time of the emitting electrons
is smaller than (or comparable to) the cascading time-scale of 
the large-scale turbulence implying that the evolution of the synchrotron 
power depends very much on the level of MHD turbulence in the ICM (e.g.,
Brunetti \& Lazarian 2010).
Consequently, when the injection of MHD turbulence
is suppressed (eg. at late
merging-phase), then also the synchrotron emission at higher 
radio frequencies is suppressed, falling below the detection limit of 
radio observations, as soon as the energy density of turbulence starts 
decreasing. 

Consequently in this scenario 
cluster {\it bi-modality} may be expected 
because the transition between radio halos and ``radio quiet'' 
clusters in the $P_{1.4}$--$L_X$ diagram is predicted to be fairly fast   
(Brunetti et al. 2007, 08) provided that the 
acceleration process we are looking in these sources is not
very efficient, being just sufficient to generate radio halos emitting 
at a few GHz frequencies (in which case radio halos disappear as soon
as a fraction of turbulence is dissipated).
Interestingly, a relatively inefficient electron acceleration process 
in radio halos (with acceleration time about 100 Myrs) is in line with 
the steep spectrum observed in some halos 
and with the presence of a spectral steepening at 
higher frequencies discovered in a few halos (e.g., Thierbach et al. 2003,
Brunetti et al. 2008, Dallacasa et al. 2009, Giovannini et al 2009).

\section{Testing turbulent acceleration with LOFAR surveys}

LOFAR promises an impressive gain of two orders of magnitude 
in sensitivity and angular resolution over present instruments in 
the frequency range 15--240 MHz, 
and as such will open up a new observational window to the 
Universe. 

The steep spectrum of radio halos makes these sources 
ideal targets for observations at low radio frequencies suggesting
that present radio telescopes can only detect the tip of the iceberg
of their population (En\ss lin \& R\"ottgering 2002; Cassano et al.~2006).

In the picture of the {\it turbulent re-acceleration} scenario, the formation 
and evolution of radio halos are tightly connected with the 
dynamics and evolution
of the hosting clusters. 
Indeed, the occurrence of radio halos at any redshift depends on 
the rate of cluster-cluster mergers and on the fraction of 
the merger energy channelled into MHD turbulence and re-acceleration 
of high energy particles. 
In the past few years, this has been modeled by Monte Carlo 
procedures (Cassano \& Brunetti 2005; Cassano et al. 2006) that 
provide predictions verifiable by future instruments.

Stochastic particle acceleration by MHD turbulence is believed
to be rather inefficient in the IGM. Consequently, electrons can 
be accelerated only up to energies of $m_e c^2 \gamma_{max} \leq$ several GeV, 
since at higher energies the radiation 
losses are efficient and hence dominate. 
The consequent spectral steepening expected in the synchrotron spectrum 
makes it difficult to detect these sources
at frequencies higher than the frequency, $\nu_s$, at which the 
steepening becomes severe;
$\nu_s$ given by (eg Cassano et al.~2010b) 

\begin{equation}
\nu_s \propto \, <B> \gamma_{max}^2 \propto {{<B> \chi^2 \,}\over{
\big( <B>^2 + B_{cmb}^2 \big)^2 }}
\label{nub}
\end{equation}

\noindent
where $\chi \simeq 4 D_{\rm pp}/p^{2}$, $p$ is the momentum of the electrons 
and $D_{pp}$ is 
the electron diffusion coefficient in the momentum space 
due to the coupling with turbulent waves. 
In the case of a single merger between a cluster with mass 
$M_v$ and a subcluster of mass $\Delta M$, Cassano \& Brunetti (2005) 
derived that $\chi$ can be approximated by

\begin{equation}
\chi \propto { { \eta_t}\over{ R_H^3 }}
\Big({ {M_{v} +\Delta M}\over{R_v}}\Big)^{3/2}
\frac{r_s^2}{\sqrt{k_B T}}
\times
\Big\{
\begin{array}{ll}
1 & {\rm if}\, r_s \leq R_H \\
(R_H/r_s)^2 & {\rm if}\, r_s > R_H\,,
\end{array}
\label{nub2}
\end{equation}

\begin{figure*}[t]
\begin{center}
\includegraphics[width=0.5\textwidth
]{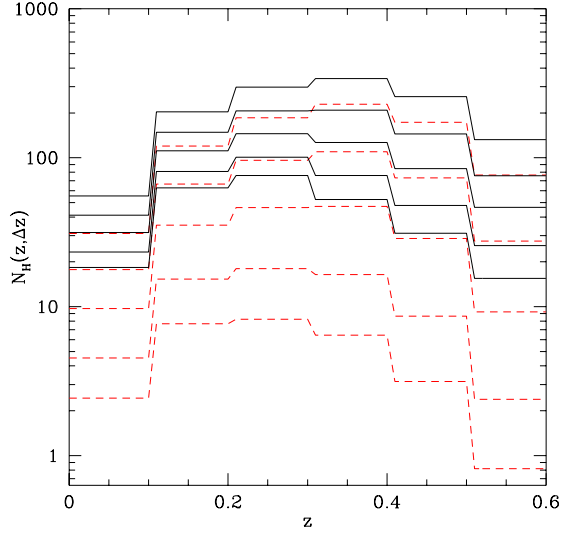}
\caption[]{All sky number of expected radio halos (with $\nu_s \geq 120$ MHz)
as a function of redshift
assuming sensitivities = 0.1, 0.25, 0.5, 1.0, 1.5 mJy/beam (solid lines,
from top to bottom) at 120 MHz (from Cassano et al 2010b).
Dashed lines show number counts for very steep spectrum
radio halos, with $\nu_s$ in the range 120--600 MHz.
} 
\label{fig:Lr_Lx}
\end{center}
\end{figure*}

\noindent
where $r_s$ is the stripping radius of the subcluster crossing 
the main cluster, i.e., the distance from the center of the subcluster 
where the static pressure equals the ram pressure 
(see Cassano \& Brunetti 2005 for details), $R_H$ is the size of the 
radio halo, and $R_v$ and $T$ are the virial radius and temperature 
of the main cluster, respectively.

\noindent
Combined with Eq.~\ref{nub}, this implies that higher values of 
$\nu_s$ are expected in the more massive clusters, 
$\nu_s \propto (M_v/R_v)^3/T \propto M_v^{4/3}$ (here considering 
for simplicity a fixed value of $B$, see Cassano et al.~2006 for a more general
discussion), and in connection with major merger events, 
$\nu_s \propto (1+\Delta M/M)^3$ 
($r_s$ in Eq.\ref{nub2} also increases with $\Delta M/M$).

\noindent
Consequently
only the most energetic merger-events 
in the Universe can generate 
giant radio halos with $\nu_s \geq$ 1 GHz 
(Cassano \& Brunetti 2005). 
Similar energetics arguments can be used to
claim that radio halos with 
lower values of $\nu_s$ must be more common, since they can be generated
in connection with less energetic phenomena, eg major mergers between
less massive systems or minor mergers in massive systems (eg
Eqs.~\ref{nub}-\ref{nub2}), that are more common in the 
Universe (eg Cassano et al.~2010b).
{\it The existence of a large number of radio halos emitting preferentially
at lower radio frequencies is a unique expectation of 
turbulent models which stems from the nature of the mechanism
of turbulent acceleration that is a poorly efficient process.}

Based on recent Monte Carlo calculations (Cassano et al 2010b),  
Fig.~3 shows the expected all-sky number 
of radio halos with $\nu_s \geq$ 120 MHz in different redshift 
intervals detectable 
by typical LOFAR surveys of different sensitivities 
($0.1$ {\ldots} $1.5$ mJy/beam, see figure caption).
The LOFAR all-sky survey, 
that should reach an rms=0.1 mJy/beam at 120 MHz,
is expected to 
detect more than 
350 radio halos at redshift $\leq$0.6, in the northern 
hemisphere ($\delta\geq0$) and at high Galactic latitudes ($|b|\geq 20$).
This will increase the statistics of radio halos by about a factor of $20$ with
respect to that produced by the NVSS.  

The spectral properties of the population of radio halos
visible by the future radio surveys at low frequencies are expected
to change with the increasing sensitivity of these surveys.
In Fig.~3
we show the total number of halos with $\nu_s \geq$ 120 MHz (solid lines) 
and the number of halos with a spectral steepening at low
frequencies, $120 \leq \nu_s \leq 600$ MHz (Cassano et al 2010b).
The latter class of radio halos has a synchrotron spectral
index $\alpha > 1.9$ in the range 250-600 MHz, and would 
become visible only at low frequencies, $\nu_o < 600$ MHz.
About 55\% of radio halos in the LOFAR all-sky survey
at 120 MHz is expected to belong to this class of ultra-steep spectrum
radio halos, while radio halos of higher $\nu_s$ are expected to
dominate the radio halos population in shallower surveys.

\section{Conclusions}

The ``GMRT Radio Halo Survey'' allows to study the statistics
of radio halos in a complete sample of X-ray luminous
galaxy clusters (Venturi et al.~2008).
These observations allows to
unveil a cluster radio {\it bi-modality} with ``radio quiet''
clusters well separated from the region of the
$P_{1.4}$--$L_X$ correlation defined by giant radio halos 
(Brunetti et al.~2007 and Figure 1a).

The connection between these radio halos and cluster mergers (Cassano et
al 2010a) suggests that 
the Mpc-scale synchrotron emission in galaxy clusters is amplified during 
mergers and then suppressed when clusters become more 
dynamically relaxed.
The separation between radio halo and ``radio quiet'' clusters in
Figure 1a, and the rarity of galaxy clusters with intermediate radio
power implies that the processes of amplification and suppression
of the synchrotron emission takes place in a relatively short
time-scale, $\tau_{evol}\approx 200$ Myr.

\noindent
At the same time the analysis of the RM of radio sources in galaxy
clusters suggests a single--mode in the clusters magnetic field
properties where radio halo and ``radio quiet'' clusters have similar
magnetic fields (Govoni et al 2010,
see also Bonafede, these proceedings).

\noindent
Clusters radio bimodality combined with the independent information
on the magnetic field in these clusters, 
provides a novel tool to constrain models 
proposed for the origin of radio halos, namely the re-acceleration 
and hadronic model.

The short transition time-scale can be potentially reconciled with the 
hypothesis that the emitting electrons are accelerated by cluster-scale 
turbulence, in which case the synchrotron radiation emitted at GHz 
frequencies may rapidly decrease as a consequence of the dissipation of 
a sizeable fraction of that turbulence.

\noindent
The most important expectation of the turbulent reacceleration scenario
is that the synchrotron spectrum of radio halos should become gradually 
steeper above a frequency, $\nu_s$, that is determined by the energetics 
of the merger events that generate the halos and by the electron radiative 
losses (e.g., Fujita et al.~2003; Cassano \& Brunetti 2005).
Consequently, the population of radio halos is predicted to consist of 
a mixture of halos with different spectra, steep-spectrum halos being 
more common
in the Universe than those with flatter spectra (e.g., Cassano et al.~2006).
The discovery of these very steep-spectrum halos will allow us to test 
the above theoretical conjectures.

\noindent
Despite the uncertainties caused by the unavoidable simplifications
in present calculations, 
about 350 giant radio halos are expected in the future LOFAR surveys.
This means that LOFAR will increase the statistics of these sources by 
a factor of $\sim 20$ with respect to present-day surveys.
About $1/2$ of
these halos are predicted with a synchrotron spectral
index $\alpha > 1.9$ and would brighten 
only at lower frequencies, which are inaccessible to present observations.
Most important, the spectral properties of the population of radio halos
are expected to change with the increasing sensitivity
of the surveys as steep spectrum radio halos
are expected to populate the low-power end of the radio halo luminosity
functions. 
The discovery of a large fraction of radio halos with spectra steeper than 
$\alpha \approx 1.5$ is expected to allow a robust discrimination between 
different models of radio halos, for instance in this case simple 
energetic arguments would exclude a secondary origin of the 
emitting electrons (e.g., Brunetti 2004; Brunetti et al. 2008).

Because of the large number of expected radio halos, 
a potential problem with these surveys is the identification of
halos and their hosting clusters.
LOFAR surveys are expected to detect radio halos 
in galaxy clusters with masses $\geq 6-7 \times 10^{14}$M$_{\odot}$ 
at intermediate redshift.
On the other hand, statistical samples of X-ray selected clusters, which
are unique tools for identifying the hosting clusters,
typically select more massive clusters at intermediate z. 
In this respect the future surveys with eROSITA and with SZ--telescopes 
will provide crucial, complementary, information.

\end{document}